\def\year{2019}\relax
\documentclass[letterpaper]{article} 
\usepackage{aaai19}  
\usepackage{times}  
\usepackage{helvet} 
\usepackage{courier}  
\usepackage[hyphens]{url}  
\usepackage{graphicx} 
\usepackage{mdframed}
\usepackage{multicol, multirow, blindtext}
\usepackage{xcolor,colortbl}
\usepackage{subcaption}
\usepackage{siunitx}
\usepackage{graphicx}  
\usepackage{amsmath}
\mdfdefinestyle{MyFrame}{
    innerrightmargin=2pt,
    innerleftmargin=2pt,
    innertopmargin=2pt,
    innerbottommargin=2pt
}

\title{What You See Is What You Get? \\
The Impact of Representation Criteria on Human Bias in Hiring}

\author{
Andi Peng, Besmira Nushi, Emre K\i c\i man, Kori Inkpen, Siddharth Suri, Ece Kamar\\
\\ Microsoft Research
\\ Redmond, WA, USA
\\ {
    \fontfamily{pcr}\selectfont 
    \{andipeng,benushi,emrek,kori,suri,eckamar\}@microsoft.com
    }
}

\begin{document}
\maketitle

\begin{abstract}
Although systematic biases in decision-making are widely documented, the ways in which they emerge from different sources is less understood. We present a controlled experimental platform to study gender bias in hiring by decoupling the effect of world distribution (the gender breakdown of candidates in a specific profession) from bias in human decision-making. We explore the effectiveness of \textit{representation criteria}, fixed proportional display of candidates, as an intervention strategy for mitigation of gender bias by conducting experiments measuring human decision-makers' rankings for who they would recommend as potential hires. Experiments across professions with varying gender proportions show that balancing gender representation in candidate slates can correct biases for some professions where the world distribution is skewed, although doing so has no impact on other professions where human persistent preferences are at play. We show that the gender of the decision-maker, complexity of the decision-making task and over- and under-representation of genders in the candidate slate can all impact the final decision. By decoupling sources of bias, we can better isolate strategies for bias mitigation in human-in-the-loop systems.
\end{abstract}

\section{Introduction}

\noindent Machine learning can aid decision-making and is used in recommendation systems that play increasingly prevalent roles in the world. We now deploy systems to help hire candidates \cite{hirevue2018}, determine who to police more \cite{veale2018fairness}, and assess the likelihood of an individual to recidivate on a crime \cite{angwin2016machine}. Because these systems are trained on real world data, they often produce biased decision outcomes in a manner that is discriminatory against underrepresented groups. Systems have been found to unfairly discriminate against defendants of color in assessing bail \cite{angwin2016machine}, incorrectly classify minority groups in facial recognition tasks \cite{raji2019actionable}, and engage in wage theft for honest workers \cite{mcinnis2016taking}. While much of the algorithmic fairness literature has focused on understanding bias from algorithms in isolation \cite{dwork2018group}, little is known about how human decisions, when interacting with these systems, are impacted \cite{green2019disparate}. 

In this paper, we study algorithmic decision-making in hiring. While there exists a long history of studying hiring discrimination in fields like economics and social psychology, there has been little work done on the interplay between these demonstrated biases and the influence of algorithmic systems, especially across a wide variation of different professions and study participants. Here, we conduct a large-scale experiment studying \textit{screening} or \textit{recommendation systems} that choose a person (or set of people) from a pool of candidates \cite{kleinberg2019discrimination}. Figure 1 shows the process flow for these hybrid systems, also known as human-in-the-loop or algorithm-in-the-loop systems. Data is gathered from the existing hiring pools in the world, often in larger quantities than what a human can feasibly assess, and fed into an algorithm which then screens for a candidate slate. The human decision-maker utilizes this filtered list to produce the final decision on whether or not a candidate should be recommended for hire. Biased gender hiring recommendations can be due to many different sources, including from (1) the world distribution (gender breakdown of candidates in a specific professions) (2) algorithmic bias in what's displayed to the human and (3) human decision-making itself.

\begin{figure}[t]
  \centering
  \includegraphics[width=.95\columnwidth]{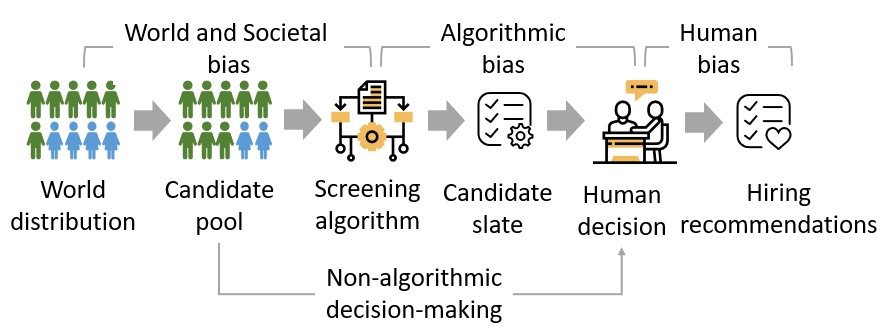}
  \caption{A high-level schematic for a hybrid system for hiring. A biased decision can be impacted by world, algorithmic, and human bias. Representation criteria is an intervention deployed when the candidate slate is generated.}
\end{figure}

\noindent {\bfseries Approach} 
This study seeks to understand the impact of different sources of bias on hiring, specifically what properties of the candidate slate impact human decision-making decisions. We define a biased decision to mean any difference in the system outcome such that one gender is favored over another in a manner that does not correspond to the gender distribution of the candidate slate as fed into the system. For example, imagine that two systems, A and B, both input the same distribution (50-50 M-F candidates). If System A produces outcomes that result in a 50-50 distribution of M-F recommended for hire but System B does not, then System B would be a biased system. We utilize the same reasoning for humans in assessing output decisions with respect to the input distribution that may come from the world (in traditional hiring practices) or from an algorithm (in hybrid systems). This view ascribes no notion of fairness or justice to the decision outcome.

We conduct experiments using Amazon Mechanical Turk to study how participants recommend candidates for different professions given candidate slates where we control factors such as education and experience, and artificially assign distributions of female and male candidates (see Figure 2). We generate profiles where we replace the names and pronouns displayed in the text and hold all other factors constant. Using representation criteria (i.e., fixed proportional display of gender distribution on the candidate slate), we randomly assign gender and ordering of profiles on each individual task to study how hiring decisions vary across different professions and gender distributions. This design affords us the following advantage: for any candidate slate, we can ask ``how would the decision outcome be different if the particular gender of the candidates were changed?''. We ask participants to rank, out of 8 total, their top 4 candidates to recommend to a friend. We vary the proportion of M-F candidates in each profession and observe the impact of representation criteria on hiring outcomes. By displaying the specific gender distributions displayed to a human decision-maker as fixed controls, we can attribute any observed disparity in hiring outcomes as bias linked to human decision-making, rather than algorithmic or world distribution biases. We compare these to baseline outcomes of the study conducted on the current world distribution of each profession and an AI model trained on word embeddings, both which represent systems that do not take into account representation criteria. We ask the following questions:

\begin{enumerate}
    \item Does balancing the gender distribution in candidate slates mitigate bias? How does this effect vary across different professions?
    \item In professions where this intervention is not enough, does over-representation help?
    \item How do personal features of the decision-maker, such as gender, impact the outcome?
\end{enumerate}

\begin{figure}[t]
  \centering
  \includegraphics[width=.95\columnwidth]{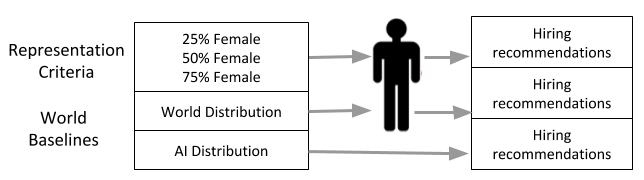}
  \caption{Our experimental design, highlighting the different candidate slates that can be displayed to the decision-maker. Representation criteria represent the intervention tested and world baselines highlight examples of existing processes.}
\end{figure}

\noindent {\bfseries Takeaways}
Our results suggest three key takeaways regarding the relationship between world distribution bias and human decision-making. First, when genders are balanced in the candidate slate, gender bias in many professions with skewed world distributions can be mitigated. However, there are some professions in which this intervention is not enough to completely correct bias. Second, for professions like nannies and OBGYNs, no adjustment of representation criteria, even to the point of extreme over-representation, can fully correct for biased outcomes, suggesting that there are {\em persistent preferences} at play with respect to which genders people prefer for specific jobs that are independent of how candidates are displayed. Third, even across the same profession, there are personal features of the decision-maker, such as gender, that impact both the direction and strength of decision bias. As we seek to understand more regarding how algorithms can be deployed safely and \textit{fairly} in the real-world, we must also study how bias from algorithms impacts decision-making by understanding the effect of different sources of bias on decision outcomes.

\section{Related Work: Sources of Bias}

\noindent {\bfseries Human Decision-Making Bias}
The opaqueness and inconsistencies of human decision-making have long been documented \cite{cunningham2013biases,glascher2011model,kahneman2003aperspective}. Although frameworks for evaluating risk \cite{rangel2009neuroeconomics}, particularly with limited information under uncertainty \cite{platt2008risky}, have been evaluated, their applicability to real-world tasks has often not been studied. Cognitive biases---unconscious, automatic influences on human decision-making that reliably produce reasoning errors---have been understood as heuristics, or mental shortcuts, that humans take when evaluating large amounts of uncertain information in a messy world~\cite{tversky1974judgment,thaler2008nudge}. These heuristics range from availability bias \cite{tversky1973availability}, our willingness to judge the frequency of events in the world by the ease with which examples come to mind, to hindsight bias \cite{fischhoff1975iknew}, our tendency to believe falsely, after the outcome is known, that we'd have accurately predicted it. While manifestations of cognitive bias have been detailed, consistent theories on bias mitigation---prevention and reduction of these biases---are still limited.

\noindent {\bfseries Gender Bias in Hiring}
Hiring has long been a discriminatory practice, with socially salient features such as race and gender identified as being impactful towards hiring outcomes \cite{isaac2009interventions,steinpreis1999impact}. We focus on the role of gender in the hiring process due to its long-standing history of inequality and current usage in real-world systems, including the latest release from the World Economic Forum which extrapolates that it will take another 118 years to close the world gender gap \cite{gendergapreport2017}. This can be due to many factors, including on average women are more likely to work part-time, be employed in low-paid jobs, and not take on management positions \cite{worldoutlook2017}. As seen in Figure 3, the world distributions for M-F candidates in different professions vary greatly, presenting a data representation problem. Furthermore, there is also evidence that gender inequalities in the workplace stem, at least in part, from bias directed against women from those who hold internal sexist attitudes or preferences for a particular gender in different professions. Studies have documented personal discrimination against women by decision-makers \cite{koch2015meta}. For instance, a study found that the higher the participants scored on a hostile sexism test, the more likely they were to recommend a male candidate rather than female for a managerial position \cite{masser2004reinforcing}. These findings demonstrate that there exist human biases that must be decoupled from world distribution problems, particularly when placed within an algorithmic process.

Moreover, while these biases have been studied directly through classic resume screening experiments \cite{bertrand2003emily} as well as indirectly through questioning techniques \cite{hoffman2019prejudice}, due to the difficulty in conducting studies assessing candidate preferences in the real world, much of this work has largely consisted of field studies focusing on isolated instances of a single profession or job call and its impact on callback rates \cite{kang2017whitened}. Algorithmic hiring and the increased usage of automated systems now motivates the need to understand this process in its entirety through decomposition of component biases, as the ability to deploy algorithms in screening processes has already been shown to have alarmingly detrimental and widespread effects \cite{amazon2018}. Furthermore, while hiring processes were often unique to the specific company or recruiter soliciting candidates, algorithmic screening now impacts individuals on a scale that was impossible to see prior. Today, platforms such as LinkedIn Recruiter or Google Hire provide a singular large-scale, centralized screening database for many professions and companies, further motivating the need for a comprehensive study of the breakdown of decision biases across a wide array of professions and decision-makers. In a centralized platform where a recruiter can utilize the same tool to screen for thousands of candidates across many different professions, how does the input distribution and representational display impact the final decision? We build on this wide expanse of work in designing experiments to better understand the integration of these factors: how do real world gender distributions across different professions integrate with algorithmic interventions to impact hiring bias in hybrid systems?

\begin{figure}[t]
  \centering  
  \includegraphics[width=.95\columnwidth]{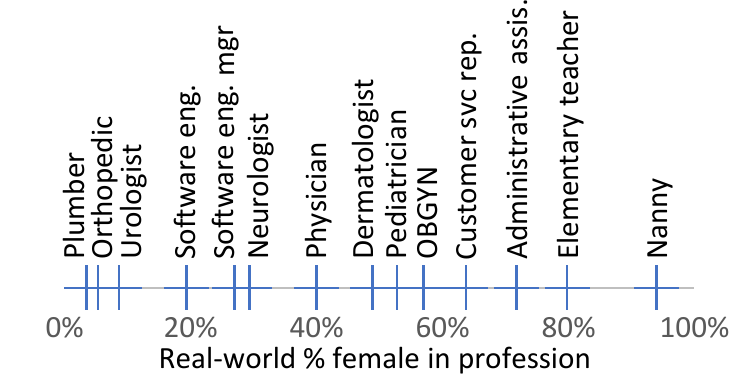}
  \caption{Real-world distributions across professions.}
\end{figure}

\noindent {\bfseries Algorithmic Bias}
Although efforts have been made to study and correct bias in algorithmic hiring \cite{raub2018bots}, replicating bias has been challenging \cite{thebault-spieker2018simulation}. Many attempts have sought to collect information on the current state of the world and, post-decision, statistically demonstrate a reduction of disparate outcomes. For example, researchers have attempted to debias a model trained on word embeddings after they found undesirable statistical links between certain jobs (such as programmer) to specific genders (such as male) \cite{bolukbasi2016man,de-arteaga2019bias}. The rise of online platforms has also resulted in discriminatory rankings of candidates, \cite{hannak2017discrimination,chen2018investigating,celis2018ranking}. For example, an empirical study found that STEM ads displayed on Facebook were targeted more at men than women \cite{lambrecht2017algorithmic}. Despite explicit attempts made by hiring platforms to improve algorithmic fairness \cite{geyik2019fairness}, we still do not understand the impact of different sources of bias and how they impact final recommendations when used by a human in the real world. Does this statistical correlation mean that people hold an internal preference for men as programmers or do the current skewed world distributions, of which AI systems only serve to accentuate \cite{zhao2017men}, create biased decision outcomes?

\section{Problem Formulation: Hiring}

\noindent {\bfseries Definitions}
We recognize that the study of bias and fairness spans many fields, including psychology, computer science, and the social sciences. We offer the following terms as a reference guide for how we employ them, understanding that they are utilized differently in other contexts.

\noindent\textit{Bias:} Any observed difference in recommendation decisions such that one gender is favored over another in a manner that does not correspond to the distribution input in to the algorithm or human. This can be impacted by algorithms, (\textit{algorithmic bias}), humans (\textit{decision-making} or \textit{human bias}), or skewed world distributions (\textit{world bias}).

\noindent\textit{Decision-making:} The human process of taking input data and outputting a decision.

\noindent \textit{Representation Criteria:} Fixed proportional display of candidates by gender. Slates can be 50-50 M-F, 75-25, etc.

\noindent\textit{World Distribution:} The gender breakdown of candidates in a specific profession. We use this as a metric for understanding the labor pools available for hire.

\noindent\textit{Candidate Slate:} The list of candidates generated from the world distribution by the screening algorithm that is displayed to a human decision-maker. Representation criteria is an intervention that alters the candidate slate.

\noindent {\bfseries World Distribution Problem}
Labor pools highlight a data representation problem, as gender breakdowns in different professions vary greatly (see Figure 3). Skews in distribution range from plumbers, where 3.5\% of the workforce is female to nannies, where 94\% are female \cite{laborforce2019}. Even within licensed MDs in the United States, specialties can range from OBGYNs, where 57\% are female to orthopedic surgeons, where 5.3\% are female \cite{activephysicians2018}. This poses a ripe opportunity to isolate gender bias in world distributions from human bias and explore how hiring decisions are impacted by unrepresentative data as opposed to human decision-making processes.

\noindent {\bfseries Algorithmic Hiring Tools}
The rise of online profiles and job platforms has led to a shift in how jobs are filled: it is now easier than ever for recruiters to find candidates and for candidates to apply for jobs without ever engaging with a human. Many processes are now, if not completely automated, supported by automated screening tools such as Google's Hire \cite{googlehire2019}.

Recent failures of algorithmic-aided hiring tools, such as Amazon's recruiting tool \cite{amazon2018}, have spurred action to create less-biased, more representative systems. One such example is LinkedIn Recruiter: a service where a human has access to a portal where they can query and filter potential candidates by job related criteria. To ensure that the top ranked candidates displayed for a search reflect the underlying talent pool with respect to gender distribution, LinkedIn deployed a ``representational ranking" algorithm \cite{geyik2019fairness}. The intervention works as follows: after ML algorithms have calculated a ``best fit" score for each candidate for a particular search, ``representation ranking" post-processes the display of candidates so that, on each page, the proportion of female candidates shown is the same as the corresponding proportion of LinkedIn profiles matching that query. Note, this does not mean a 50-50 split is shown---rather, the goal is to reduce any gender skew that the AI system may have generated. However, the resulting effectiveness of such approaches has not been studied.

\noindent \textbf{Research Question 1:} Does balancing the gender distribution in candidate slates mitigate bias? How does this effect vary across different professions?

\noindent \textbf{Research Question 2:} For professions where this intervention is not enough, does over-representation help?

\noindent \textbf{Research Question 3:} How do personal features of the decision-maker, such as gender, impact human decision-making in hiring recommendations?

\section{Human-in-the-Loop Methodology}
We conduct large-scale experiments to study the effect of representation criteria on gender bias. We generate textual profiles for different professions and randomize the assignment of gender to individual candidates. We then ask over 4,900 unique study participants to select their top candidates to recommend for hire. Across trials, we vary representation criteria: the proportion of female and male candidates displayed. We compare these results against the expected distribution based on two world baselines: the current world representation for that profession and an AI score generated by projecting each profession across the gender subspace in a word embedding \cite{mikolav2013linguistic}.

\noindent {\bfseries Profile Generation}
We study the following professions: doctors (dermatologists, neurologists, OBGYNs, orthopedic surgeons, pediatricians, physicians, and urologists), nannies, plumbers, administrative assistants, software engineers, software engineering managers, customer service representatives, and elementary teachers. For each profession, eight fictional candidates are generated using online biographies as inspiration. Our experimental data underwent ethics, privacy, and IRB review. All profiles follow roughly the same setup and include: name, degree, specialty or focus within the profession, experience length (shown explicitly through years of work or implicitly via graduation year), and information on personal life. This was done to induce low variance and control for possible confounding factors such as education and expertise. To further ensure profile equivalency, we run pilot studies asking participants to differentiate between candidates and remove profiles that were ranked too many or few times. We then randomize the gender assignment of each candidate with respect to the slate tested. For each profile, the number of times the name and gendered pronouns appear is kept constant. We generate our profiles using the most common female and male first names according to the 2018 Census, keeping last names constant. Textio analysis is run on each profile to remove other gendered words \cite{textio2015}. All profiles are consistently 300-400 words long.

    \begin{figure}[t]
        \centering
        \tiny
        \begin{tabular}{|p{0.95\linewidth}|}
        \hline
             Dr. \textbf{Robert} Brown, MD, is a board-certified orthopedic surgeon who, since 2002, practices at the Cleveland Clinic in Beachwood, OH. \textbf{He} is a graduate of the Johns Hopkins School of Medicine and completed \textbf{his} residency in Cleveland. \textbf{He} spends much of \textbf{his} time educating medical students at the Cleveland Clinic Lerner College of Medicine of Case Western Reserve University, where \textbf{he} serves as a Orthopedics Advisor and as Course Director for rotations that integrate bone fracture prevention and healthy living. \textbf{His} practice interests include health maintenance and diet/exercise, in addition to joint replacement. In \textbf{his} free time, \textbf{Robert} enjoys biking and exploring the outdoors.  \\
             \\
             Dr. \textbf{Mary} Brown, MD, is a board-certified orthopedic surgeon who, since 2002, practices at the Cleveland Clinic in Beachwood, OH. \textbf{She} is a graduate of the Johns Hopkins School of Medicine and completed \textbf{her} residency in Cleveland. \textbf{She} spends much of \textbf{her} time educating medical students at the Cleveland Clinic Lerner College of Medicine of Case Western Reserve University, where \textbf{she} serves as a Orthopedics Advisor and as Course Director for rotations that integrate bone fracture prevention and healthy living. \textbf{Her} practice interests include health maintenance and diet/exercise, in addition to joint replacement. In \textbf{her} free time, \textbf{Mary} enjoys biking and exploring the outdoors.\\
        \hline
        \end{tabular}
        \caption{Example of a male and female candidate assignment for an orthopedic surgeon profile.}
    \end{figure}

\noindent {\bfseries HIT Generation}
Using these profiles, we generate HIT tasks where each participant is shown a unique candidate slate and asked to recommend candidates to \emph{a friend who is looking for a specific profession}. This is done with the intention that a recommendations to a third-party would induce less personal bias. Each HIT has the same layout, where participants are shown 8 candidates and asked to rank, from 1 to 4 (with 1 as their top choice), their top candidates out of 8. We enforce representation criteria by randomly assigning gender to each profile according to the distribution studied (for example, a 25F slate would randomly assign 2 female and 6 male candidates, then randomly order their display). All gender assignments vary across individual HITs in a condition, with no participant ever shown the same slate. This is done to remove possible confounding factors such as rank ordering preferences, recency bias, and variance across profiles to isolate the gender variable. Afterwards, participants are asked to fill out a survey regarding their demographic information and to qualitatively describe which factors they believe to be important in their decision. Each HIT requires 5-10 minutes to complete.

\noindent {\bfseries Ranking Task}
We deploy our experiments using a crowdsourced pool on Amazon Mechanical Turk and compensate workers at a wage of \$15 per hour. All participants are preliminarily screened according to the following qualifications: hold above a 90\% approval rating, unique study participant, and based in the United States. For each profession, 300 unique study participants are recruited and each assigned a candidate slate of 8 profiles. The experiment is deployed in 3 batches for each profession: 100 HITs for gender slates of 25\% female-75\% male (25F), 50\% female-50\% male (50F), and 75\% female-25\% male (75F). We exclude duplicate or missing ranks. All tasks were completed between January 18 and April 19, 2019.


\noindent {\bfseries Statistical Testing}
A hypergeometric distribution, which represents unbiased hiring decisions, is generated for each representation criteria (25F, 50F, 75F) to model the expected number of female candidates selected in the top 4 (out of 8). Such a distribution models the discrete probability distribution of binary draws \textit{without} replacement from a finite population. This is relevant to our task as the probability of each candidate being ranked changes on each draw. Note that, modeling the screening decision-making process with a hypergeometric random variable $X \sim$ hypergeometric$(K,\, N,\, n)$ depends on the total number of candidates ($N$), the number of candidates from a gender group ($K$), and the number of candidates to be selected ($n$). We intentionally remove other factors (such as expertise), to study the effect of representation criteria in isolation. Here, $\text{Pr}(X=k)$ denotes the probability that exactly $k$ candidates from the candidate slate made it onto the list of $n$ recommended candidates. An unbiased decision process that does not favor any one gender should follow (as closely as possible) a hypergeometric distribution with the following probability mass function:

$$ \text{Pr}(X=k) = \frac{{K\choose k} {N - K\choose n - k}}{{N\choose n}} $$
If the decision process does not follow such a distribution, then the decision-making entity is either (i) biased towards a specific group ; e.g., towards female candidates if the mass of the observational data distribution is concentrated above the hypergeometric mean (i.e., $\frac{nK}{N}$), or (ii) there exist other factors that govern the decision process leading to decision bias. Since we randomly assign gender to profiles for each HIT and control for equal expertise in the profile text, deviations of the observational distribution from the corresponding hypergeometric can only be attributed to gender bias.






\begin{figure}[t]
  \centering
  \includegraphics[width=.95\columnwidth]{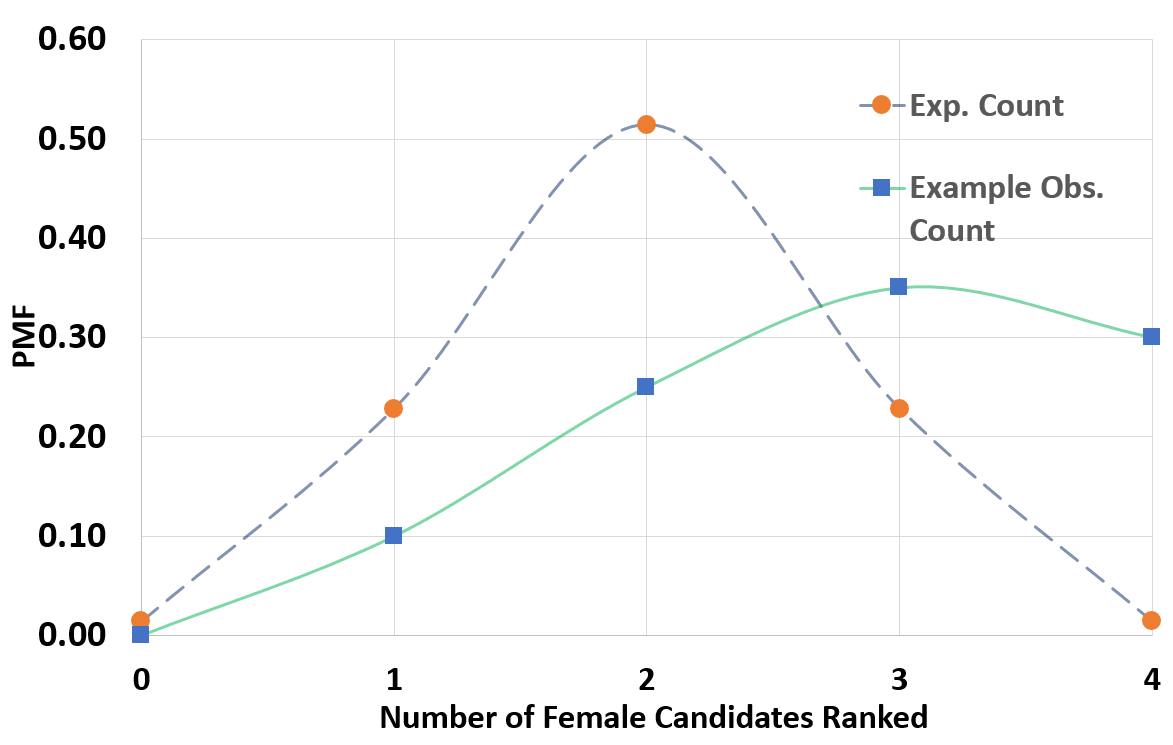}
  \caption{Example probability mass function for hypergeometric $(K=400,\, N=800,\, n=400)$ where, for 100 ranking slates where each slate displays 4/8 candidates as women (50F), the expected number of total female candidates ranked out of 4 follows this distribution. The dashed line represents the ideal and the solid line a biased recommendation distribution in favor of women.}
\end{figure}

We conduct Chi-square goodness-of-fit tests for our observed distributions against the hypergeometric to test if selections of female candidates deviate significantly. Note: the hypergeometric captures the number of female candidates selected, but not the ranking order. We study ranking by reformulating the problem as recommending 1 candidate out of 8, 2 out of 8, etc and observing the distribution of female candidates selected. The closer the observed distribution of rankings is to the expected, the higher the correlation value and consequently lower the observed gender bias is.

\noindent {\bfseries World Baselines}
To determine the cross-sectional impact of representation criteria on hiring outcomes, we create two baselines: current world representation and an AI score generated from word embeddings. They are given as additional representation criteria to model potential hiring decisions that are currently utilized today by decision-makers. We recognize that these do not fully represent real-world AI hiring systems, but are useful for understanding the potential impact of real world data on algorithmic systems.

\noindent {\bfseries Current World Representation}
To address the discretization challenge of displaying world distributions that were not multiples of 8 (for example, it is impossible to display 3\% of plumbing candidates out of 8 as female for one slate), we run the required number of candidates to create the appropriate proportion across all HITs. For example, we ran 24 HITs of 12.5F for plumbers to create 24 female candidates out of the 800 total candidates required for $p_F = 0.03$ and $n = 100$. We then test our observed rankings against a weighted hypergeometric distribution of the same proportions.

\noindent {\bfseries AI Score}
Hiring platforms do not publicly share their ranking algorithms, so to better model the impact of world distributions on algorithmic systems, we create example AI decision scores by projecting each profession across a gender direction as represented by word embeddings trained on a Google News text-corpora. A word embedding is a representation of a single word (or common phrase) $w$ as a $d$-dimensional word vector $\overrightarrow{w} \in R^d$ \cite{mikolav2013linguistic}. Word embeddings, trained on word co-occurrence in a mass text-corpora, can serve as a dictionary of sorts to analyze word meaning for words with similar meaning tend to have vectors that are closer together. To understand the gender biases present across professions (which professions are closer to $\textit{she}$ than to $\textit{he}$), we utilize the w2vNEWS embedding, a $d$ = 300-dimensional word2vec pre-trained on the Google News corpus, and project the profession of interest across the gender direction $v_{g}$ \cite{bolukbasi2016man}. The cosine similarity between each word vector $w^d$ and the gender direction $v_g$ is represented by the following equation:

\begin{center}
$
cos(w^d, v_g) = \frac{w^d \cdot v_g}{||w|| ||v_g||}
$
\end{center}

We then convert the similarity into a gender score \cite{bolukbasi2016man} by normalizing the resulting correlation against the same $\overrightarrow{\textit{she}}-\overrightarrow{\textit{he}}$ vector to obtain a score from $0-100$, where $0$ represents a perfect \textit{she} match, $50$ a gender-neutral word, and $100$ a perfect $\textit{he}$ match.

\section{Results}

\begin{enumerate}
    \setlength\itemsep{0em}
    \item Balancing gender slates can serve as an effective intervention for mitigating bias across many professions, even those that have extremely skewed world distributions.
    \item For professions where this is not effective, the human decision-making bias is so strongly in favor of a specific gender that it dominates any over-representation.
    \item Personal features of the decision-maker, such as gender, can also impact the hiring outcome.
\end{enumerate}

\begin{table*}[h]
\footnotesize
\begin{tabular}
    {|p{2.6cm}||p{1.15cm}|p{1.15cm}|p{1.15cm}|p{1.15cm}|p{1.15cm}|p{1.15cm}|p{1.15cm}|p{1.3cm}|p{1.3cm}|}
    \hline
     \textbf{Profession} & \textbf{0F Selected} & \textbf{1F Selected} & \textbf{2F Selected} & \textbf{3F Selected} & \textbf{4F Selected} & \textbf{\%F World} & \textbf{AI Score (0-100)} & \textbf{\%F Rank4 ($p$)} & \textbf{\%F Rank1 ($p$)}\\
     \hline
     \textit{\textbf{Exp. count ($n$=100)}}  & \textbf{1.4} & \textbf{22.9} & \textbf{51.4} & \textbf{22.9} & \textbf{1.4} & 50.0 & 50.0 & 50 & 50\\
     Plumber & 2 & 22 & 55 & 18 & 3 & 3.5 & 40.0 & 50(0.513) & 46(0.424)\\
     Ortho. surgeon & 2 & 26 & 57 & 12 & 3 & 5.3 & 31.6 & 47(0.086) & 54(0.428)\\
     Urologist & 6 & 23 & 48 & 22 & 1 & 8.7 & 38.0 & \cellcolor{yellow}47(0.005) & 53( 0.549)\\
     Software eng. & 0 & 20 & 53 & 24 & 3 & 19.3 & 36.5 & 53(0.460) & 47(0.549)\\
     Software eng. man. & 1 & 24 & 57 & 17 & 1 & 27.0 & 40.6 & 48(0.659) & 48(0.689)\\
     Neurologist & 2 & 22 & 56 & 17 & 3 & 29.4 & 42.9 & 49(0.420) & 51(0.842)\\
     Physician & 1 & 23 & 48 & 26 & 2 & 40.0 & 43.1 & 51(0.907) & 51(0.841)\\
     Dermatologist & 5 & 28 & 49 & 18 & 0 & 48.9 & 65.0 & \cellcolor{yellow}45(0.013) & 45(0.317)\\
     Pediatrician & 2 & 27 & 40 & 28 & 3 & 52.8 & 66.2 & 51(0.171) & 56(0.230)\\
     OBGYN & 1 & 16 & 40 & 27 & 16 & 57.0 & 74.2 & \cellcolor{yellow}60(\textless 0.000) & 59(0.072)\\
     Customer serv. rep. & 0 & 29 & 51 & 20 & 0 & 63.7 & 65.9 & 48(0.301) & 50(1.000)\\
     Admin. assistant & 1 & 17 & 49 & 31 & 2 & 71.7 & 59.0 & 54(0.301) & 59(0.072)\\
     Elem. teacher & 1 & 15 & 56 & 27 & 1 & 79.8 & 67.9 & 5 (0.391) & 54(0.424)\\
     Nanny & 0 & 18 & 40 & 35 & 7 & 94.0 & 78.2 & \cellcolor{yellow}58(\textless 0.000) & 53(0.549)\\
    \hline
\end{tabular}
 \caption{Observed number of female candidates ranked in the top 4 by profession for the 50F distribution. The top row details the expected hypergeometric values. The percentage of female candidates recommended for hire for the top 4 and top 1 ranking tasks are compared to the current world representation and normalized AI scores calculated from word embeddings. The $p$-value represents the Chi-square test of the distribution of observed female candidates selected against the expected hypergeometric.}
\end{table*}

\subsection{Bias Induced by the World}

We show findings from experiments where we present a gender balanced candidate slate for all professions to test the effectiveness of this as an intervention strategy. We discuss the impact of candidate gender in observed outcomes by comparing the observed ranking of female candidates to their expected distributions to decouple the impact of human decision-making bias from world bias.

\begin{figure*}[t]
   \centering
   \includegraphics[width=.95\columnwidth]{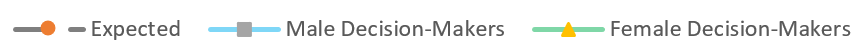}
\begin{tabular}{lcccc}
\parbox[t]{2mm}{\multirow{2}{*}{\rotatebox[origin=c]{90}{\small\fontfamily{pcr}\selectfont{PMF}}}}
{\small(a)}\includegraphics[width=3.6cm]{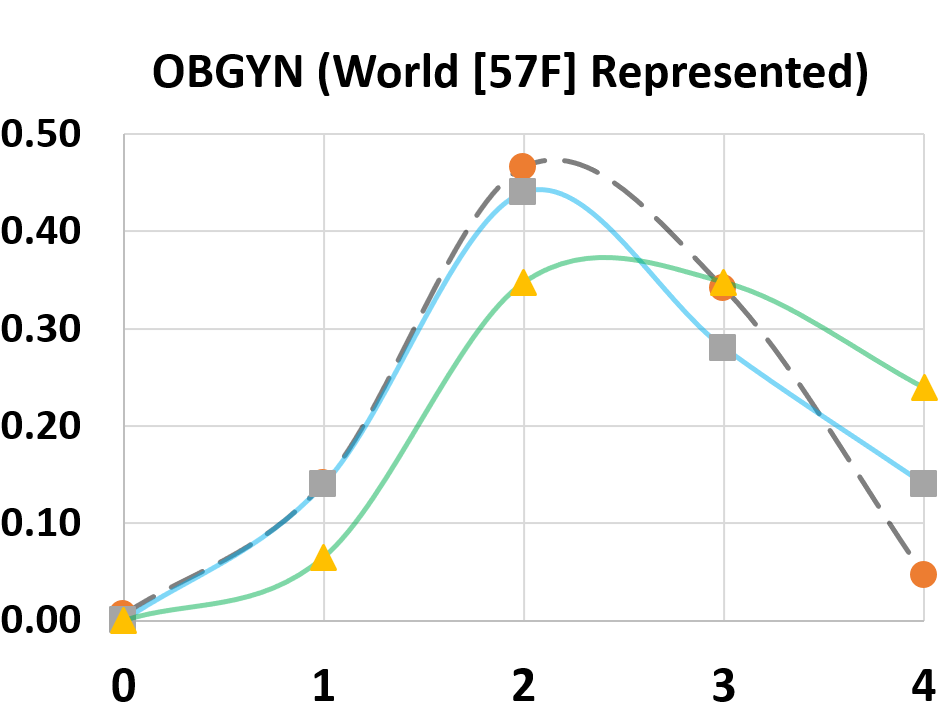}&
{\small(b)}\includegraphics[width=3.6cm]{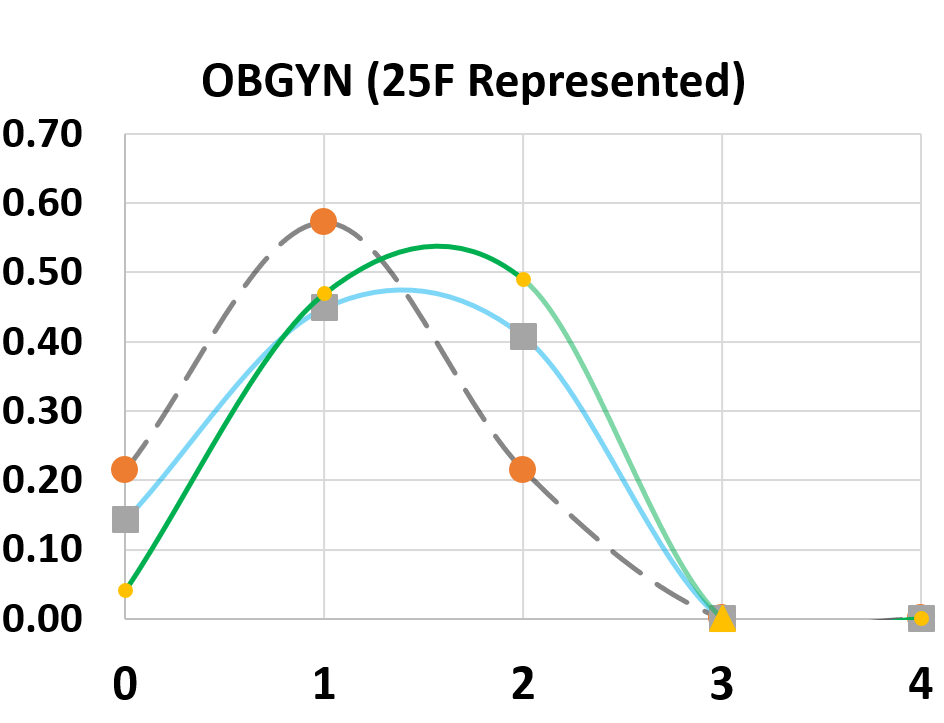}&
{\small(c)}\includegraphics[width=3.6cm]{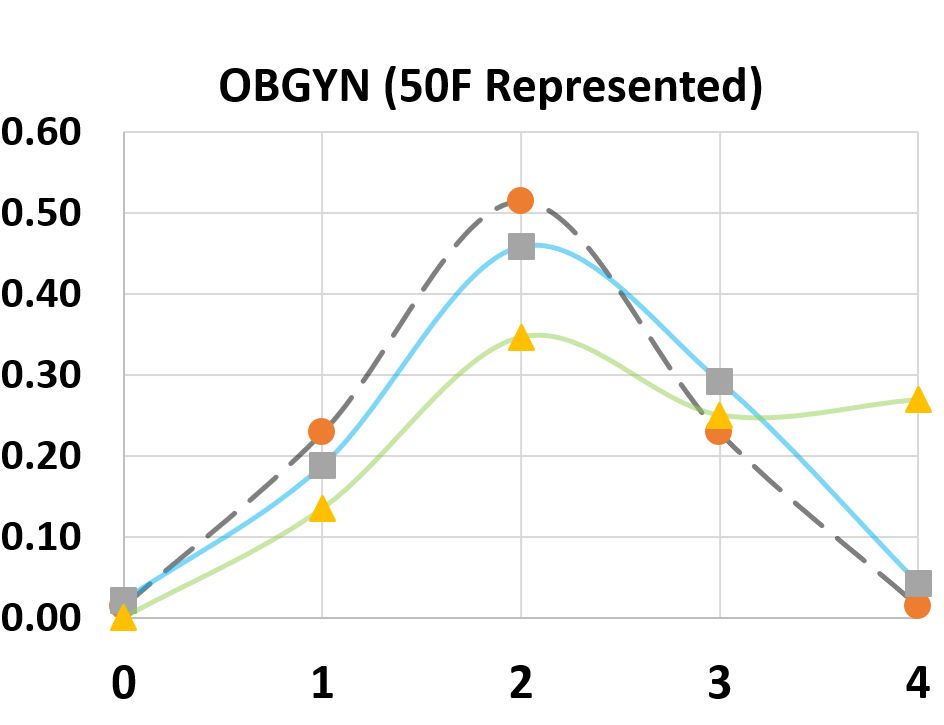}&
{\small(d)}\includegraphics[width=3.6cm]{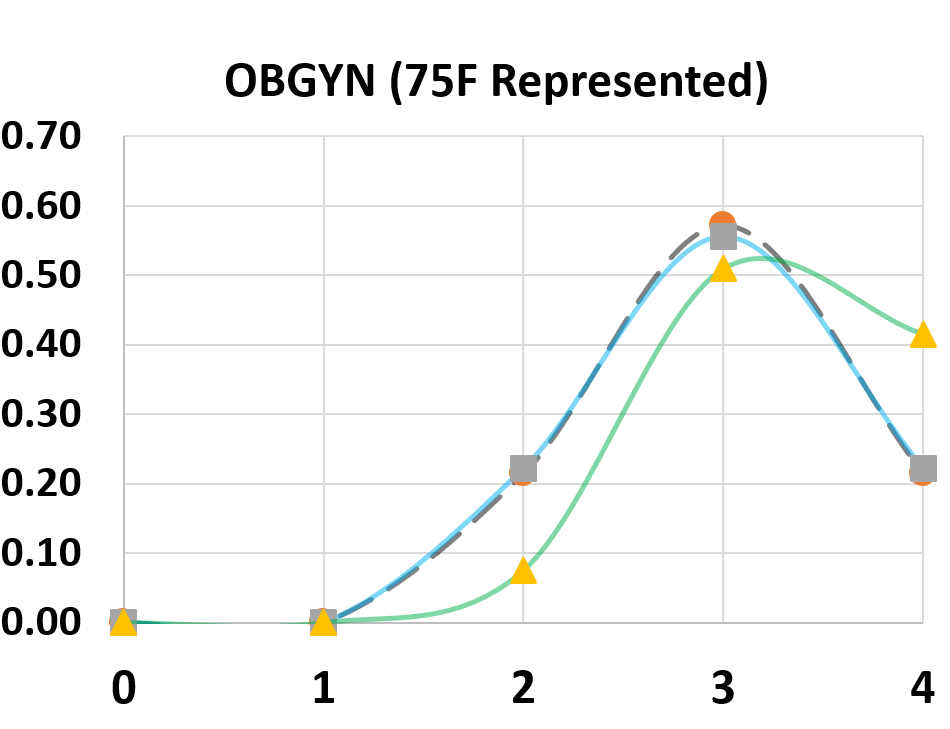}\\
{\small(e)}\includegraphics[width=3.6cm]{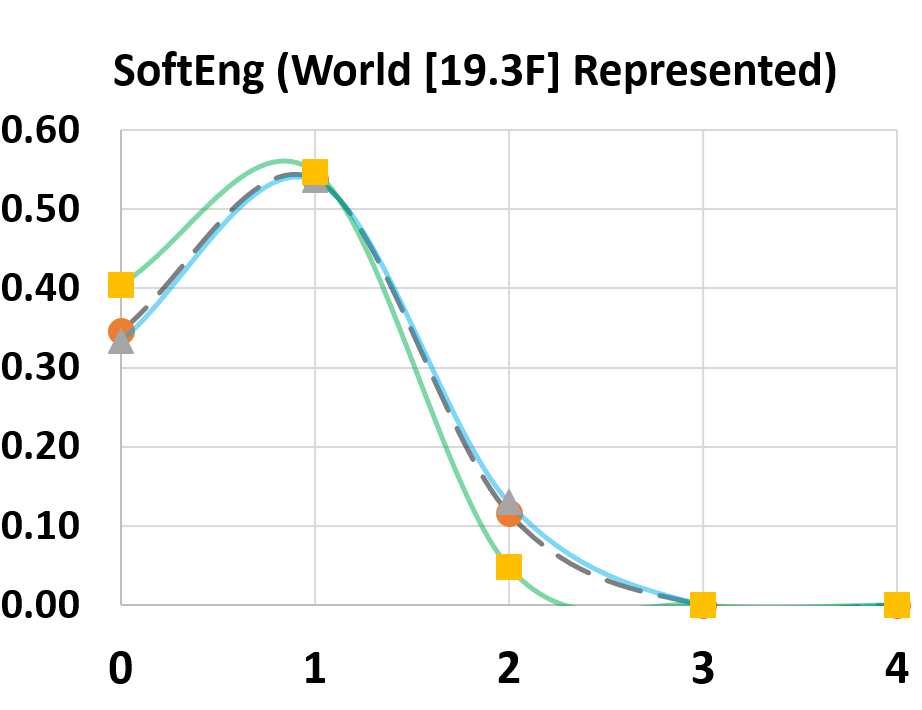}&
{\small(f)}\includegraphics[width=3.6cm]{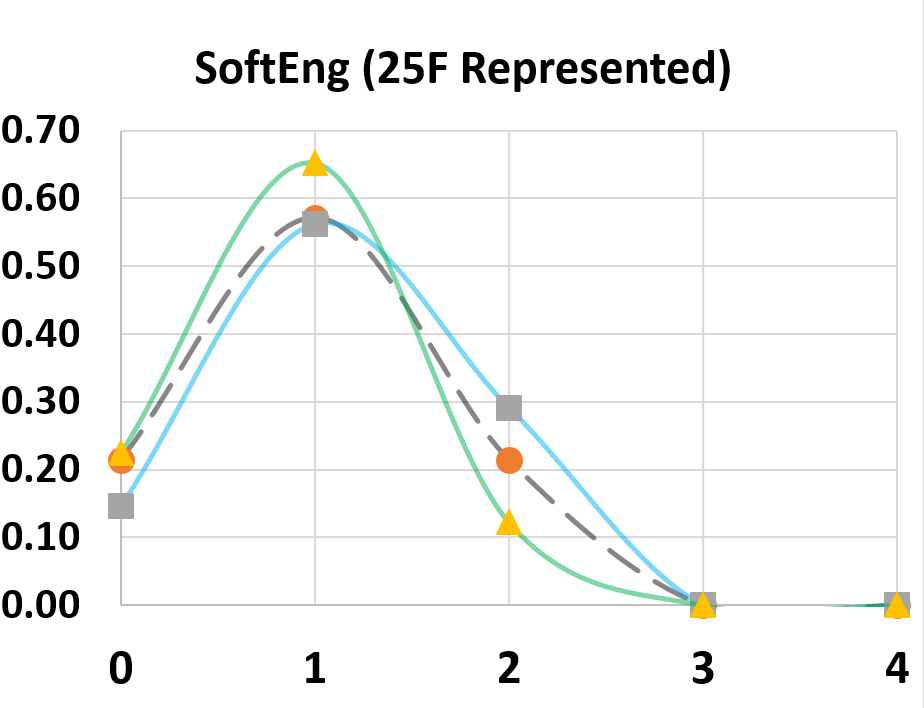}&
{\small(g)}\includegraphics[width=3.6cm]{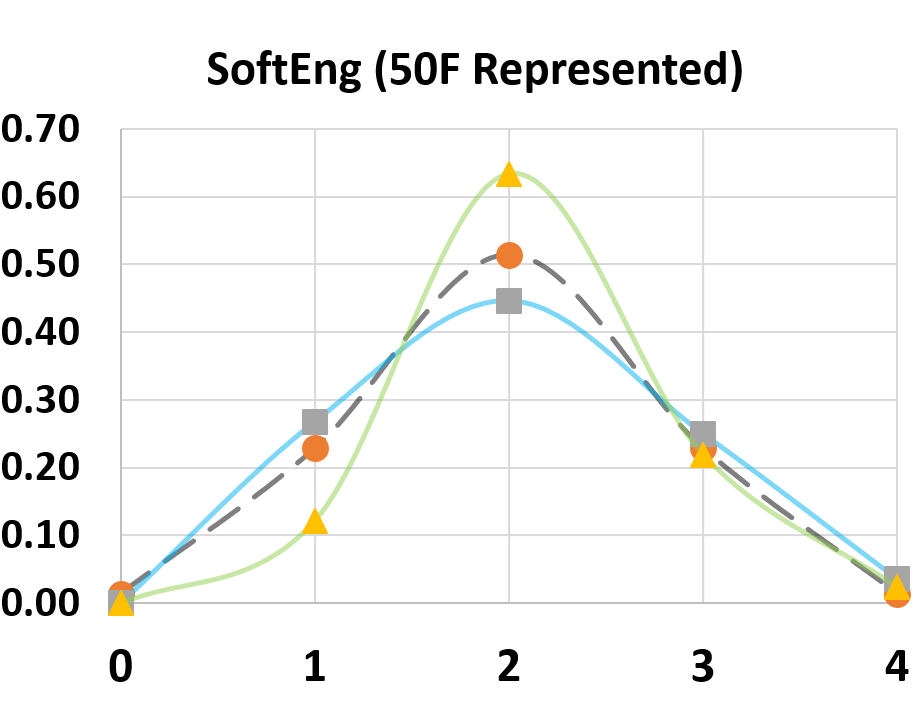}&
{\small(h)}\includegraphics[width=3.6cm]{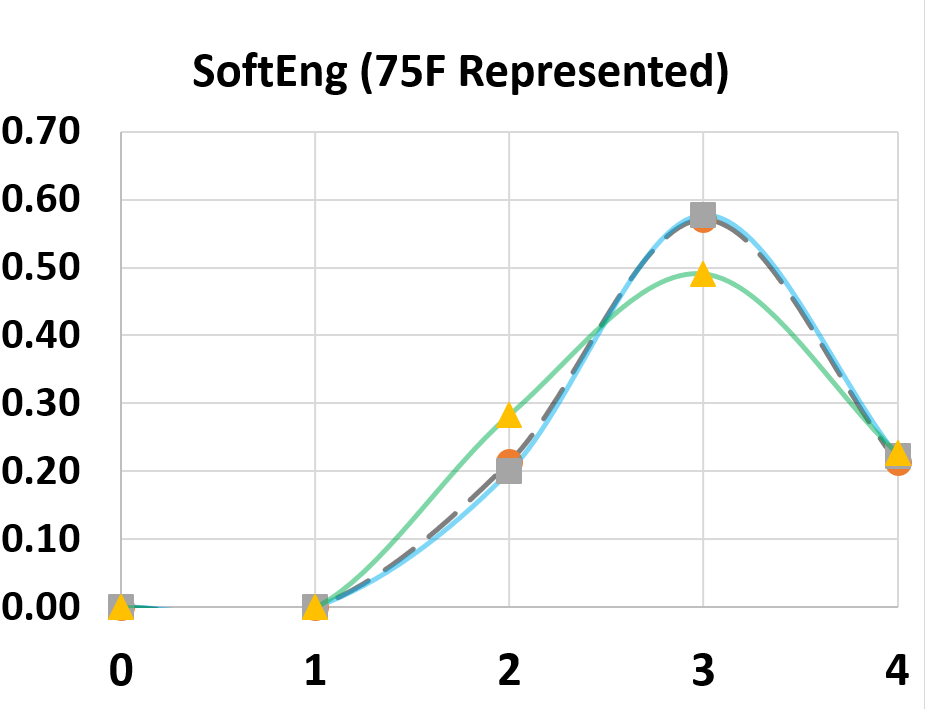}\\
\multicolumn{4}{c}{\small\fontfamily{pcr}\selectfont{Number of Female Candidates Ranked}} \\
\end{tabular}
    \caption{Hypergeometric distributions of the expected and observed rankings for OBGYNs and software engineers by gender of the decision-maker. We observe near-universal bias in favor of female OBGYNs but none for software engineers.}
\end{figure*}

\noindent {\bfseries Gender of Candidate}
Hiring decisions do not not always reflect the world distribution. When we control the gender representation of candidate slates to be equal (50F), many  outcomes can be seen as unbiased and follow the expected hypergeometric distribution, not the input representation.

\begin{mdframed}[style=MyFrame]
\textbf{Result 1:} Balancing candidate slates to equally represent gender can serve as an effective intervention for bias mitigation across many professions.
\end{mdframed}
Despite many professions where the world distribution is radically skewed towards one gender, we see no bias if gender is represented equally across a candidate slate. For example (see Table 1), although only 5.3\% of orthopedic surgeons in the world are female, we observe no human bias in the gender balanced ranking tasks (\%F Rank4 and \%F Rank1). This was also seen in plumbers, software engineers, software engineering managers, neurologists, physicians, pediatricians, customer service representatives, administrative assistants, and elementary teachers. This finding suggests that some unrepresentative professions may be entirely due to the lack of candidates in the hiring pool and not from inherent human biases, and creating a balanced slate can serve as an effective form of bias mitigation.

\noindent \textbf{Observation 1a:} Slates that are not gender balanced can create biased decisions. Male participants exhibit bias against female candidates in ranking dermatologists ($\chi$ = 6.13, 0.047) and urologists ($\chi$ = 9.39, 0.009) when candidate slates are 75F on the 4-rank task. This suggests a desire to over-represent male candidates when the distribution does not favor them. On the other hand, male participants exhibit bias against male candidates in ranking OBGYNs ($\chi$ = 11.05, 0.004) and physicians ($\chi$ = 6.76, 0.034) when candidate slates are 25F on the 4-rank task, meaning that they wish to over-represent female candidates when few are present. Female participants also exhibit bias against female candidates in ranking urologists ($\chi$ = 10.67, 0.005) when candidates slates are 25F on the 4-rank task.

These results suggest that for some professions, balancing the genders might not illicit biased decisions but unbalanced representations in favor of one gender does. For example, male participants may hold an internal preference for wishing to see male dermatologists and urologists, professions where there could exist a sense of relatability in understanding concerns, but that preference only emerges when the candidates slates are skewed against male candidates.

\noindent \textbf{Observation 1b:} Some professions never create biased decisions, regardless of the representation criteria.
We observe several professions such as neurologists, pediatricians, plumbers, administrative assistants, software engineers, software engineering managers, and customer service representatives, where 25F and 75F skewed slates show no bias in hiring decisions, i.e., the resulting distributions reflect the expected distributions for these slates (see Figures 6e-g). We postulate this is because in these professions, there is no clear reason why one gender would be better suited to succeed at the job, and the lack of bias generated suggests that representation criteria remain a valid intervention strategy for skewed world distributions.

\subsection{Bias Induced by Humans}
For professions where balancing the candidate slate was not effective in mitigating bias, we test over-representation of a specific gender. We show findings from experiments where we present gender inbalanced slates for all professions to study the strength of human decision-making bias. For each profession, we test candidates slates of 25F, 50F, and 75F and real-world baselines (representation criteria generated from world distributions and example AI scores).

\begin{mdframed}[style=MyFrame]
\textbf{Result 2:} Despite over-representation of candidates, some professions consistently produce biased decisions, suggesting there are persistent human biases at play as to which gender is preferred for certain professions.
\end{mdframed}
Some professions demonstrate persistent human biases that are not impacted by representation criteria at all. For all candidate slates ranging from 25F to 75F, female decision-makers demonstrate a clear and persistent preference for female candidates and over-represent them on all ranking tasks and representation distributions (see Figures 6a-d).

For example (see Table 1), the number of female OBGYNs ranked out of 8 was far more than expected, with 16 participants ranking all 4 selected candidates vs. the expected 1.4 (4F Selected). This was also seen in urologists, dermatologists, and nannies, suggesting that persistent preferences for who humans prefer for certain jobs may be ingrained in some professions, possibly due to societal stereotypes of what gender is best suited for specific jobs or personal comfort related to the individual. These could not be mitigated through any representation or display criteria. This motivates future work in discovering other important factors that take into account cultural and societal expectations.

\begin{figure}[t]
  \includegraphics[width=.95\columnwidth]{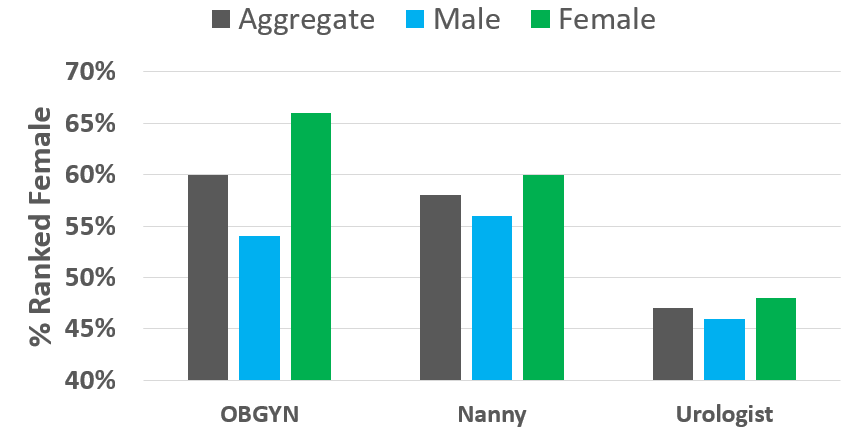}
  \caption{Professions where female decision-makers primarily drove the overall biased outcomes.}
\end{figure}

\noindent {\bfseries Gender of Decision-Maker}
We decompose our data across the identity of the decision-maker. Figures 7-8 detail how human bias can be separated by decision-making gender.

\begin{mdframed}[style=MyFrame]
\textbf{Result 3:} There are personal features of the decision-maker, such as gender, that impact hiring outcomes. Sometimes a biased outcome is driven primarily by one gender and other times an unbiased outcome is created due to opposing effects from each gender that cancel out.
\end{mdframed}

\noindent \textbf{Observation 3a:} Biased hiring decisions are at times driven by strong effects from one decision-making gender.
For a subset of the professions where we observe bias, the aggregate effect is primarily driven by one decision-making gender. Figure 7 shows that strong bias in professions like OBGYNs and nannies is mostly attributed to female decision-makers. Here, the weaker or nonexistent effect by male participants is subsumed by the stronger effect demonstrated by female participants. This suggests that for certain jobs, there exist personal preferences that are ingrained within the identity of the individual decision-maker. 

\noindent \textbf{Observation 3b:} Aggregate unbiased decisions are sometimes hidden by opposite biases from decision-making genders. 
For a subset of the professions where no aggregate hiring bias is seen, we find differences when we split the data by the gender of the decision-maker. Figure 8 shows how opposite effects from male and female participants create an aggregate unbiased decision for some professions. This suggests that bias decomposition into factors related to the decision-maker's identity is important, as signals can be lost when looking at aggregate outcomes alone.

\subsection{Bias Induced by the Task}
We now detail additional observations related to how the complexity and display of the task may have impacted the outcomes presented. We show findings from differences in hiring outcomes across participants' top 1, 2, and 4-ranked candidates and qualitative feedback collected.

\begin{figure}[t]
  \includegraphics[width=.95\columnwidth]{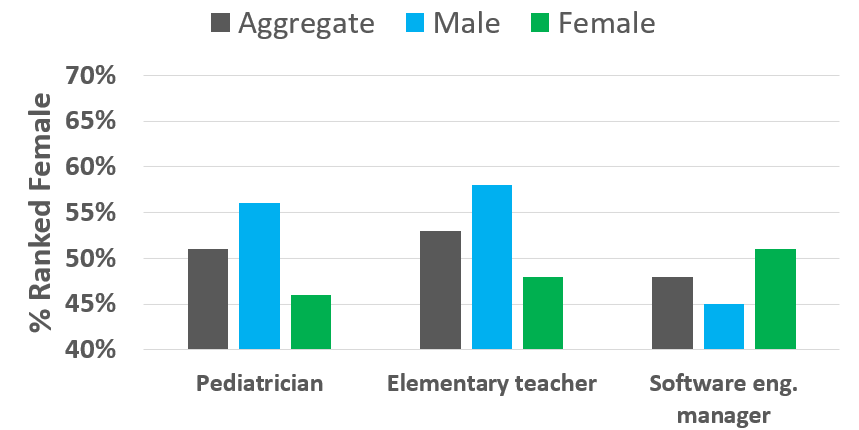}
  \caption{Professions where unbiased outcomes are cancelled out by opposing decision-making genders.}
\end{figure}

\noindent \textbf{Observation 4a:} Decision-makers rarely demonstrate bias for the top candidate, but exhibit bias in later rankings.
As seen in Table 2, in professions where aggregate biased decisions is observed in top 4 rankings (Human Select 4), the effect decreases for the top 2 ranking (Human Select 2) and disappears entirely for the top 1 ranking (Human Select 1). This may be due to task fatigue, where humans, when asked to rank one candidate, focus on more objective factors related to competency such as education and experience. However, when the task becomes more complex, markers for the best candidates disappear and cognitive biases take over. This suggests that interventions can be tailored to specific tasks depending on the complexity and amount of information that decision-makers are exposed to.

\noindent \textbf{Observation 4b:} Participants are more willing to express their choice of a female candidate as being important to their ranking decision. For OBGYNs, the strongest bias seen in any profession tested, both male and female participants overwhelmingly state that they expected a friend to prefer a female doctor for addressing feminine needs. However, when we test urologists, the male counterpart, we do not observe the same willingness to comment on a preference for male doctors. We postulate this is due to an understanding of societal norms surrounding when it is acceptable to express preference, and that overt preference for a female candidate in under-represented fields is considered more \textit{fair}.

\noindent \textbf{Observation 4c:} Participants often give justifications for decisions that do not correlate to the effect observed, particularly when representations are unbalanced. For example, while we observe significant male bias for male urologists and dermatologists on the 75F task, almost none of the justifications indicate that gender is influential on the decision. Instead, factors such as ``experience" and ``education" are stated, even though these are controlled variables and random assignment of gender and order to each profile for each HIT eliminates the possibility that biases can be linked to factors other than gender. This may be due to the phenomenon that humans often have a difficult time justifying their decisions (or were not aware of their own biases) and instead look retroactively for explanations to rationalize their choices \cite{kleinberg2019discrimination}.

\begin{table}[t]
\footnotesize
\begin{tabular}{ |p{1.8cm}||p{1.7cm}|p{1.7cm}|p{1.6cm}|}
 \hline
 \bfseries{Profession} & \bfseries{Human } & \bfseries{Human} & \bfseries{Human}\\
  & \bfseries{Select 4} & \bfseries{Select 2} & \bfseries{Select 1}\\
 \hline
 Urologist & \cellcolor{yellow}${\chi}^2$ = 15.02 (0.005) & ${\chi}^2$ = 0.78 (0.678) & ${\chi}^2$ = 0.36 (0.549)\\
 Dermatologist & \cellcolor{yellow}${\chi}^2$ = 12.66 (0.013) & ${\chi}^2$ = 0.87 (0.647) & ${\chi}^2$ = 1.00 (0.317)\\
 OBGYN & \cellcolor{yellow}${\chi}^2$ = 154.10 ($p$ $<$ 0.000) & \cellcolor{yellow}${\chi}^2$ = 20.66 \xspace\xspace\xspace\xspace($p$ $<$ 0.000) & ${\chi}^2$ = 3.24 (0.072)\\
 Nanny & \cellcolor{yellow}${\chi}^2$ = 33.18 \xspace\xspace\xspace\xspace($p$ $<$ 0.000) & \cellcolor{yellow}${\chi}^2$ = 18.51 \xspace\xspace\xspace\xspace($p$ $<$ 0.000) & ${\chi}^2$ = 0.36 (0.549)\\
 \hline
\end{tabular}
 \caption{Chi-Square test statistics and p-values for top 1, 2, and 4-ranked candidates.}
\end{table}

\section{Discussion}
This project seeks to understand how different types of bias in hybrid systems impact final decision outcomes. While bias is defined in a pragmatic statistical manner in this paper, we emphasize the need to integrate social and political notions into broader discussions on algorithmic fairness.

\noindent \textbf{Experimental Validity}
As with all mTurk experiments, there are concerns of experimental validity. First, hiring is a complex task. Although we detail a breakdown of the workflow governing how a system for screening candidates might work, we undoubtedly failed to capture other important variables, including other forms of bias (e.g., insider referrals of privileged candidates). Second, while we pre-screen for high-quality mTurk workers and exclude results where participants spend too little time, there is still error from participants who did not afford the task their undivided attention. Third, we do not study a real AI system that algorithmically recommends candidates. Future work that studies bias in real world decisions and investigates other confounding factors should be explored to complement these results (e.g. how much more expertise must a female software engineer have to be selected equally frequently in a screening process?). It is also important to recognize that some of our interventions would not be possible in the real world because of extreme availability skews of certain genders in the hiring pool. For example, we generate candidate slates that are 50\% female for plumbers even though the world distribution is 3\% female. This remains a pipeline problem. Future work that studies more nuanced and realistic representation criteria is needed.

\noindent \textbf{Data Segregation}
We do not study the impact of representation criteria at the individual level but rather as aggregate groups. It is important to know how collective groups behave, but this does not give us a granular level understanding of how representation criteria impacts individual decision-making. As seen in Observations 3a and 3b, important signals can be lost when studying aggregate outcomes. We slice the data by the gender of the decision-maker, but recognize there are many other personal features that can be analyzed such as age, race, and occupation. Future work is needed to study other confounding factors. Informed segregation of data is important if different forms of intervention will be needed for different groups of decision-makers or even individuals who might exhibit unique biases.

\noindent \textbf{Choosing for Self vs. Others}
We ask participants to recommend candidates for a friend with the intention of creating a separation between the personal preferences of the human vs. review of the qualifications of each candidate. However, the processes that govern if an individual were to select for themselves may differ and should be separately assessed.

\noindent \textbf{Binary Gender}
We only study male and female candidates. The challenges for analyses of non-binary gender are two-fold. First, in studying the personal features of the decision-maker, we were unable to achieve the requisite number of study participants that would identify as non-binary. Second, it is not clear how to assign a non-binary gender to profiles that are textual, for the signal of ``they" may be interpreted in multiple ways, such as a data anonymization strategy. We recognize the troubling trend present in the algorithmic fairness literature on the lack of available work studying gender as a non-binary variable and hope that future work integrates this consideration.

\section{Conclusion}
Our study begins to decompose biases in decision-making by separating effects of world bias from the human decision-making process itself. Experiments across professions with varying gender proportions show that balancing the candidate list can correct for some professions where the world distribution is extremely skewed, although it has no impact on other professions where persistent preferences are at play. We show that the gender of the decision-maker, complexity of the decision-making task and over- and under-representation of genders in the candidate list can have confounding effects on final decisions. We believe that such a method of study, where bias mitigation is not merely examined from an algorithmic perspective in isolation but also from a hybrid system perspective, is of utmost importance as we continue to integrate automated decision systems within existing human processes. We hope this work serves as a foundation for a more nuanced understanding of bias and how to build systems for effective mitigation.

\section{Acknowledgements}
We would like to thank Adam Kalai for his wisdom on word embeddings, Krishnaram Kenthapadi for feedback on algorithmic hiring tools, Mary Gray and the MSR ethics and privacy team for IRB and standards support, and Jacki O'Neill, Luke Stark, and Ben Fish for helpful discussions.

\bibliographystyle{aaai.bst}
\bibliography{citations.bib}
\end{document}